\documentclass[11pt,letterpaper]{article}

\usepackage{fullpage}
\usepackage[margin=1.1in]{geometry}
\usepackage{times}
\usepackage{graphicx}
\usepackage{hyperref}
\usepackage{booktabs}
\usepackage{enumitem}

\hypersetup{colorlinks=true, linkcolor=blue, urlcolor=blue, citecolor=blue}

\title{Data-Oriented Modeling for Spacecraft Design}
\author{Nathan Strange\\
Vis Viva Space\\
Ventura, CA\\
}

\begin{document}

\maketitle

\begin{abstract}
Spacecraft development costs remain high despite falling launch costs, in part because
Model-Based Systems Engineering (MBSE) tools carry the complexity of the object-oriented
programming paradigm: tightly coupled data and logic, mutable state, and rigid class
hierarchies that resist integration with discipline-specific analysis tools.
This paper presents a data-oriented approach to MBSE that adapts the
Entity-Component-System (ECS) architecture from the video game industry.
Design data is stored as immutable, format-agnostic components in a generic data
system; stateless analysis functions operate on this data through templates and
containerized tools within a continuous integration pipeline.
A prototype implementation, VVERDAD (\url{https://github.com/VisVivaSpace/vverdad-prototype}), demonstrates the approach on an example
interplanetary mission concept, showing how data-oriented
principles can reduce deployment complexity, simplify testing,
and preserve the traceability benefits of document-based systems engineering.
\end{abstract}

\newpage

%----------------------------------------------------------------------
\section{Introduction} 

The space industry faces a critical bottleneck: while launch costs have plummeted,
spacecraft development remains expensive due to slow and inefficient design processes.  Complex designs exacerbate this issue by increasing
the likelihood of costly failures when attempting to cut costs.

Digital Engineering (DE) and Model Based Systems Engineering (MBSE) approaches are a
potential solution to this design complexity crisis but currently face three key challenges 1) the deployment complexity
and training difficulty for projects adopting DE/MBSE,
2) limited availability of discipline-focused engineering subsystem and science models that integrate with DE/MBSE systems, and 3) limited
DE/MBSE tool compatibility with existing engineering and science tools.

This paper presents a new approach to DE/MBSE that
adapts software technologies from
the video game industry to address these shortcomings.  A prototype implementation, VVERDAD (Vis Viva Engineering, Review, Design,
and Architecture Database), demonstrates this approach on an example mission concept.

\section{The Problem: Object-Oriented MBSE} 

Traditional Model-Based Systems Engineering (MBSE) state of the practice has several shortcomings that arise from the heavy
influence of the object-oriented programming paradigm on MBSE modeling and data management approaches. Specifically:
1) storing model data in objects with the model logic makes maintenance difficult, 2) functions with internal state make testing difficult, and
3) complex class-based data models makes extension difficult.
 
An object-oriented MBSE approach might, for example, define a class hierarchy of data file types where
each format implements its own loading and rendering logic.  This creates
code duplication (each format implements the same rendering logic),
hidden state (mutable fields in file objects),
tight coupling (rendering logic tied to file type),
and testing complexity (need to mock file I/O).

These shortcomings are compounded in aerospace engineering by several domain-specific challenges:

\begin{itemize}
  \item \textit{Heterogeneous Data Formats:} Vehicle design data comes from diverse sources --- propulsion data in YAML from supplier databases,
    mission parameters in JSON from trajectory optimization tools, power budgets in Excel spreadsheets from electrical engineers,
    and orbital elements from astrodynamics simulations.  An object-oriented approach requires a different class for each format.

  \item \textit{Data Provenance and Traceability:} Safety-critical aerospace systems require tracing every data point to its source.
    Mutable object graphs obscure this provenance.

  \item \textit{Auditability:} Design reviews need to answer: ``What data produced this analysis result?''  Hidden state in objects
    makes the data flow difficult to trace.

  \item \textit{Parallelism:} Large vehicle designs benefit from parallel processing (e.g., analyzing multiple subsystems concurrently).
    Mutable shared state in objects requires locks and careful synchronization.
\end{itemize}

%----------------------------------------------------------------------
\section{Data-Oriented Programming for MBSE} 

Parts of the video game industry have adopted data-oriented programming and functional programming paradigms
to address these shortcomings of the object-oriented approach.  Table~\ref{tab:dop} shows how these
approaches can be adapted to DE/MBSE:

\begin{table}[ht]
\centering
\caption{The Data-Oriented and Functional Programming Paradigms adapted to MBSE}
\label{tab:dop}
\begin{tabular}{p{0.45\textwidth} p{0.45\textwidth}}
\toprule
\textbf{Data-Oriented Principle} & \textbf{MBSE Benefit} \\
\midrule
\textit{Separate Data from Code:} Program logic is separated from data.  Data and functions are not bundled together in objects.
& Model and analysis code is independent from design and mission data, facilitating code reuse and reducing test complexity. \\
\addlinespace
\textit{Generic Data Interface:}  The same interface is used to access all data, and there is no need to call different functions for different types of data.
& Standard generic data interfaces make it easier to expand model functionality without worrying about breaking object and class relationships. \\
\addlinespace
\textit{Immutable Data Structures:}  Data cannot be changed.  Instead new versions are created.
& Immutable data simplifies version control, concurrency, and asynchronous modeling. \\
\addlinespace
\textit{Stateless Functions:} Functions have no state and will always produce the same output from the same input data.
& Stateless functions only need to be tested once as there is no worry about tests being invalidated in edge cases caused by unexpected program state. \\
\bottomrule
\end{tabular}
\end{table}
 
\noindent {\bf Principle 1: Separate Code from Data:}
Code (behavior) lives in standalone functions.  Data lives in plain structures with no methods.  In the VVERDAD prototype,
data components such as file contents and directory structure are pure data containers with no business logic.
Logic for loading, transforming, and querying data lives in standalone functions that receive data as parameters.\\

\noindent {\bf Principle 2: Represent Data with Generic Structures:}
All design data, regardless of source format (JSON, YAML, TOML, RON, CSV, XLSX), can be represented with
the same small set of data structures: scalars, maps, sequences, and domain-specific types (physical quantities, time epochs, tables).
This enables format-agnostic processing: template rendering operates on these generic data structures without knowing the source format.\\

\noindent {\bf Principle 3: Data is Immutable:}
Transformations create new data rather than mutating existing data.  This ensures data provenance tracking,
safe concurrent access, easier debugging and testing, and clear data flow.  In VVERDAD, user input data is never modified by analysis outputs ---
source data always takes precedence.\\

\noindent {\bf Principle 4: Separate Data Schema from Data Representation:}
Schema validation is separate from data representation.  Multiple formats can share the same validation logic.
VVERDAD does not enforce a schema at load time.  Instead, data is loaded permissively into generic data structures, and
templates reference expected keys.  This separation allows loading partial or evolving datasets, flexible data schemas
across different aerospace vehicle types, and schema evolution without code changes.
 
%\subsection{Comparison with Object-Oriented Approaches}

%Table~\ref{tab:oop-dop} summarizes the key differences between the object-oriented and data-oriented approaches for MBSE:

%\begin{table}[ht]
%\centering
%\caption{Object-Oriented vs.\ Data-Oriented Approaches to MBSE}
%\label{tab:oop-dop}
%\begin{tabular}{p{0.45\textwidth} p{0.45\textwidth}}
%\toprule
%\textbf{Object-Oriented MBSE} & \textbf{Data-Oriented MBSE} \\
%\midrule
%Data and methods in classes & Pure data in components, logic in functions \\
%Custom classes per data format & Generic data structures for all formats \\
%Mutable object graphs & Immutable data transformations \\
%Inheritance hierarchies & Enum variants and pattern matching \\
%Mock objects and DI frameworks for testing & Pure functions, no mocking needed \\
%Locks and mutexes for concurrency & Immutable data with automatic parallelism \\
%\bottomrule
%\end{tabular}
%\end{table}

%----------------------------------------------------------------------
\section{Entity-Component-System Architecture} 

VVERDAD uses a data-oriented modeling architecture based upon the Entity-Component-System (ECS) data-oriented
architecture.  ECS is used in the video game industry to facilitate software development with distributed, interdisciplinary teams
of programmers and designers.  The ECS software architecture is also well suited to parallel programming and leveraging modern graphics cards
for efficient rendering and evaluation of game physics.

\begin{figure}[ht]
\centering
\includegraphics[width=0.84\textwidth]{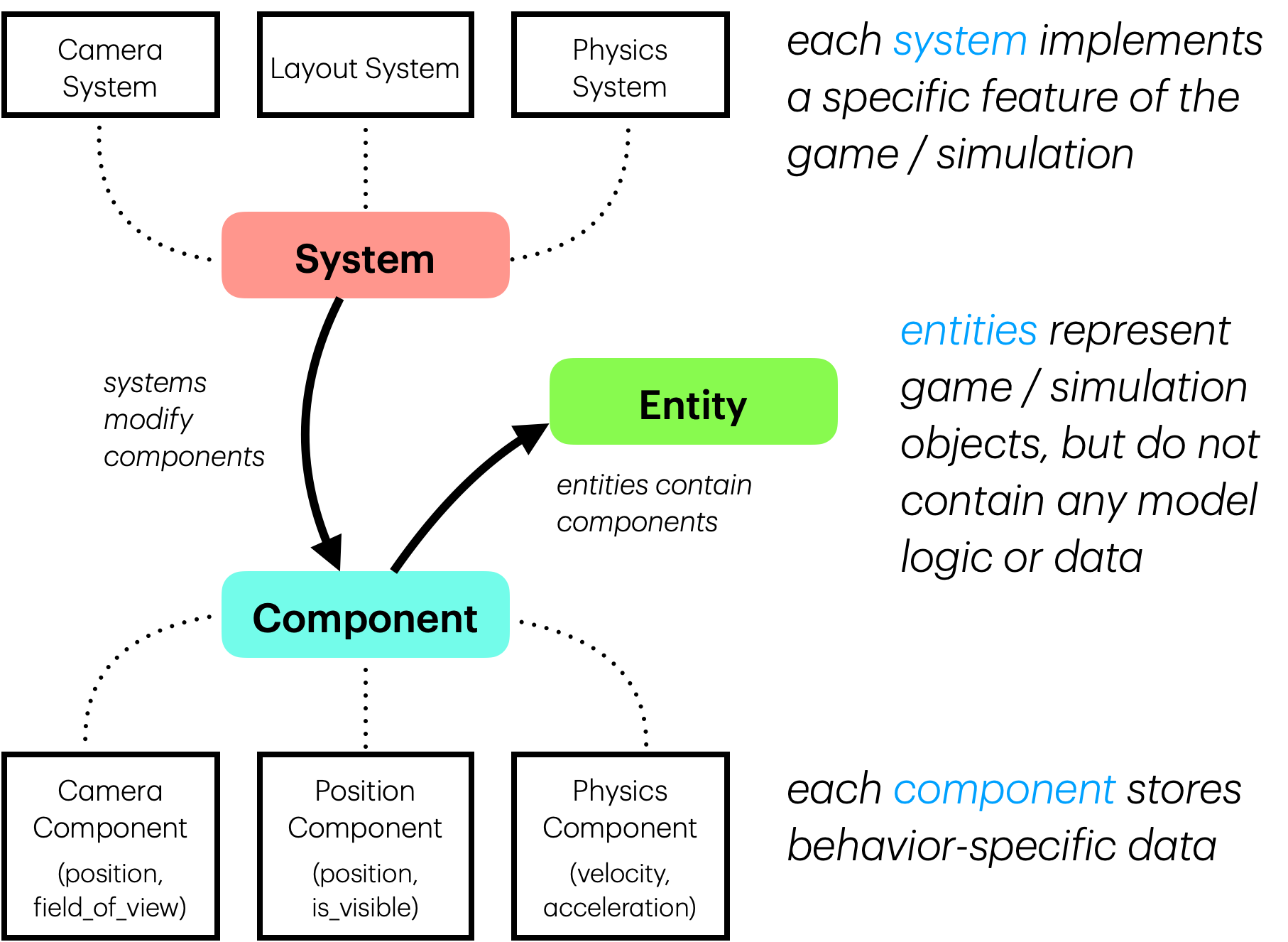}
\caption{Video Game Entity-Component-System (ECS) Software Architecture}
\label{fig:ecs}
\end{figure}

Figure~\ref{fig:ecs} presents an ECS architecture from a notional video game.  Video game simulations are structurally similar to
engineering simulations, making adaptation of the ECS architecture to engineering straightforward.  
For data-oriented MBSE, the \textit{entities} would group design
data in structures convenient for engineering calculations. The \textit{components}  would group entities to
represent design configurations, trade study options, mission scenarios, etc. The \textit{systems} would then
operate on components to perform engineering calculations, adding new data entities for the results of
these calculations.

As an analogy to traditional engineering methods: think of components as individual parameters in the design data, entities as
input files collecting the design data, and systems as engineering analysis programs processing these input files to create
output files with the results of this analysis.  With ECS, we don't need to spend effort figuring out the best way
to structure the design data before we run the analysis code.  Instead, we structure the data with a focus on understanding using
design reports, tables, and charts after the analyses are run.

In application to aerospace design software, the ECS architecture allows a simple key-value database to store entities and
components without requiring a detailed ontology to define the system components and interactions.
Instead, a more flexible namespace based on the project's directory structure can be used to access data needed for system and 
sub-system modeling.  This generic data approach frees the
MBSE team to iterate meta-model designs of system elements and relationships without delaying or
disrupting sub-system analyses.  In addition, treating the component data as immutable with a versioning system that tracks all changes
facilitates parallel execution of analysis models and model testing, and also allows traceability of design changes.

The ``systems'' in ECS can perform engineering model analyses in parallel as part of a Continuous Integration / Continuous
Development (CI/CD) pipeline that updates analyses and tests as design data is updated. These ECS systems can be implemented
as scripts generated with templates from the ECS database, allowing easy integration of existing engineering analysis tools.

This data modeling approach allows a gentler learning curve for discipline engineers: to use the tools,
they only need to learn one generic system for representing their engineering data and
integrate domain-specific analysis models into the system.  Engineers don't need to learn a proprietary data model --- they only
need to organize their data files in standard formats (JSON, YAML, TOML, CSV) in a way that makes sense to them.
The template system then allows each discipline to translate data from other teams into their own system.

\section{Preserving the Benefits of Document-Based Systems Engineering}

Prior to the development of MBSE, aerospace systems engineering was traditionally document-based.   Although this
system was less agile and slower to iterate than modern MBSE systems, the document-based approach had several important
benefits:  writing long-form prose forces engineers to clarify their thinking, documentation provides traceability of
design decisions and analysis which helps track down the root cause of problems, and documents facilitate independent
review by helping outside experts to understand the design and analysis.

The VVERDAD data-oriented MBSE approach facilitates the development of software tools that preserve
the benefits of document-based systems engineering with
data-oriented MBSE. Engineering documents can be generated as `views' into the engineering data.   These
documents can contain plots and tables generated dynamically from the analysis data with open source literate programming systems
such as Jupyter notebooks, Quarto, or Pluto.   This approach allows expanding upon the benefits of document-based systems
engineering with automated consistency checks and even documentation-generated testing from modern software engineering
practice.
 
VVERDAD also provides a data annotation system that allows review comments, questions, issues, and suggestions to be attached to
specific data points without modifying the original data files.  These annotations are stored as sidecar files alongside
the data, preserving the document-based engineering review process while
adding the traceability and searchability benefits of a structured data system.

 \section{VVERDAD: A Prototype Implementation} 

The VVERDAD prototype demonstrates this data-oriented MBSE approach as a working software tool.  VVERDAD processes a project
directory containing design data files and Jinja2-compatible templates.  It loads all data into a Bevy ECS database,
renders templates with that data, and optionally executes analysis scripts in Docker containers.  Results feed back into the database
for downstream templates.

\begin{figure}[ht]
\centering
\includegraphics[width=\textwidth]{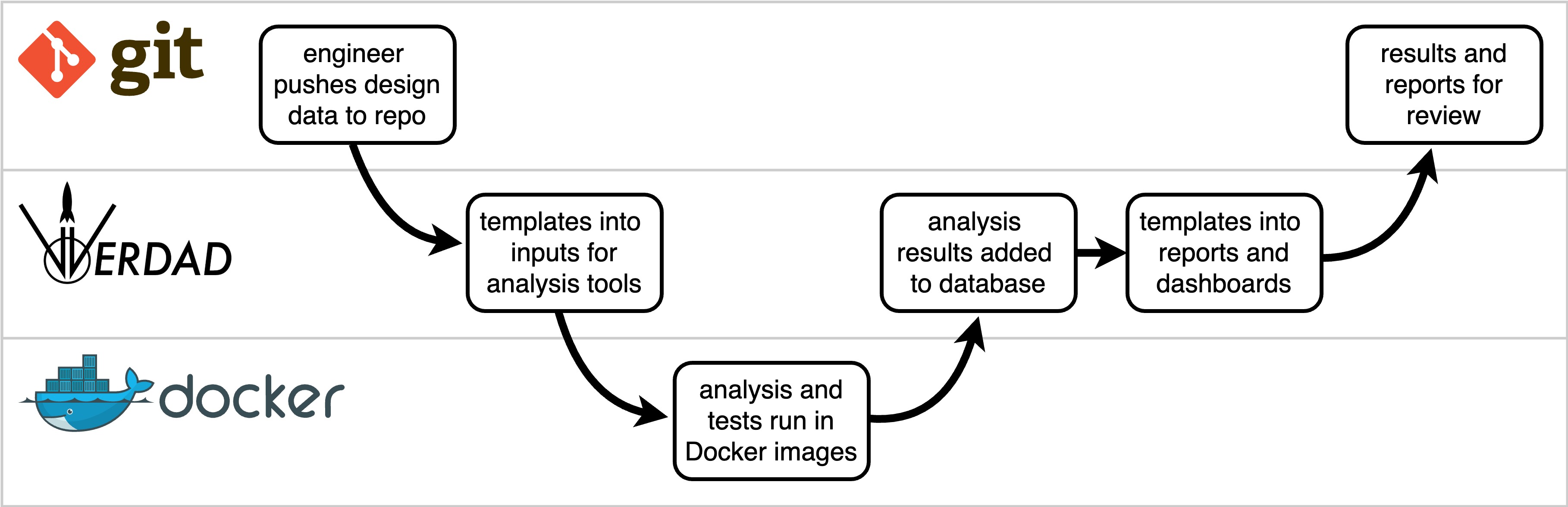}
\caption{VVERDAD Continuous Integration Workflow}
\label{fig:workflow}
\end{figure}

Figure~\ref{fig:workflow} shows how the VVERDAD core works with other open source tools such as git and Docker as part of a continuous integration pipeline.
Engineers and scientists update design data on a git repository.
Standard git repository CI/CD pipelines (GitHub Actions, GitLab CI, git hooks)
 trigger VVERDAD automation that
uses template files to generate input files for analysis tools.  These analysis tools then run in Docker images.
VVERDAD processes the outputs from the Docker-containerized analyses, adding analysis
data to its database with the design data.  VVERDAD also generates reports, dashboards, and other visualizations from template files.
Finally, all of VVERDAD's data and other products are added to the git repo for review by engineering and science teams as part of
a continuous development process.

\subsection{Implementation Language}
 
The VVERDAD core engine is developed in the \textit{Rust} programming language using the open source \textit{Bevy ECS} system~\cite{bevy}. In 2024,
the US White House Office of the National Cyber Director recommended memory-safe languages like Rust for future software
development to reduce the potential vulnerabilities to cyber attack~\cite{oncd}. As a distributed system with proprietary and ITAR-controlled
design data this is an important consideration for VVERDAD.  In addition to its memory-safety benefits, Rust's robust error checking approach
leads to lower software maintenance costs (although this is at the expense of additional up-front development complexity).

The Bevy ECS system is the module that implements ECS for the Rust language \textit{Bevy} video game engine.  The Bevy ECS system allows a low-latency,
generic data interface with parallel CPU and GPU calculation.

\subsection{Data Infrastructure}
 
 VVERDAD's data interfaces are implemented using the Rust \textit{serde} package~\cite{serde}, a framework for
efficiently and generically serializing and deserializing Rust data structures.  It allows programs to convert data
into various text and binary data formats such as CSV, JSON, YAML, TOML, BSON, MessagePack, etc.

All input formats normalize to a single generic data type with 14 variants covering scalars (float, integer, boolean, string),
physical quantities (a numeric value with a physical unit), time epochs (UTC and TDB, stored as days after J2000.0),
tabular data (from CSV and Excel files), markdown content (with optional YAML front matter),
provenance metadata, annotations, and compound types (maps and sequences).

This format-agnostic design means that a sub-system configuration can be stored in JSON, TOML, or YAML without changing the
validation or processing logic.  The data normalization pipeline follows a simple path: file bytes are parsed by a
format-specific parser into an intermediate representation, which is then converted into the generic data type with
eager parsing of domain-specific types.
VVERDAD natively supports unit and time system conversions when parsing data files and templates, eliminating 
these types of data translation errors.

\subsubsection{Project Directory as Data Namespace}
 
A VVERDAD project is a directory tree containing data files and templates.  The directory hierarchy maps directly to template
key namespacing: directory and file names become dot-separated keys.  For example, a file at
\texttt{propulsion/engine.yaml} containing a \texttt{thrust} field is accessed in templates as \texttt{propulsion.engine.thrust}.
This structure enables organizing complex designs by subsystem while maintaining clear data provenance
and template access patterns.

\subsection{Template System}
 
VVERDAD uses the \textit{MiniJinja} template engine~\cite{minijinja}, a Rust implementation of the Jinja2 template language.
Templates are recognized by their file extension (\texttt{.j2}, \texttt{.jinja}, etc.) and can be placed anywhere in the project directory tree.
The template extension is stripped when generating output files (e.g., \texttt{report.md.j2} renders to \texttt{report.md}).

Templates have full access to the project's data tree using dot notation, and can use custom filters for
unit conversion, time system operations, and data formatting.  This allows engineering reports
to be generated dynamically from the design data with plots, tables, and calculations
populated from the analysis results.
 
The MiniJinja template system was selected because its data model can be inspected at run time.  This allows VVERDAD to check
template dependencies before filling them out and to flag incomplete templates as well as to trigger
additional analyses to compute derived quantities if needed.

\subsection{Analysis Integration}
 
Analysis bundles are self-contained analysis packages identified by a \texttt{.analysis/} directory suffix.
Each bundle contains a manifest file declaring the analysis name, Docker image, input dependencies, and expected outputs,
along with template files that are rendered before execution and static files that are copied to the output directory.
This design allows VVERDAD to integrate with many different kinds of engineering analysis software with minimal friction.

VVERDAD discovers analysis bundles, renders their templates with the full project data, executes them in Docker containers,
and validates that the expected output files were produced.  Docker is optional --- if unavailable, analysis bundles
are discovered and rendered but not executed, allowing template-only workflows without Docker infrastructure.

\subsection{CI/CD Integration}
 
VVERDAD is a stateless CLI tool designed for automation.  It reads input, produces output, and requires no persistent state between runs.
A scaffolding command generates CI/CD configuration files for GitHub Actions workflows, GitLab CI/CD pipelines,
and git pre-commit and pre-push hooks.  This allows design teams to validate templates on every commit and run
full analysis pipelines on every push, catching integration errors before they propagate.

\section{Conclusion}

This paper presents a data-oriented Model-Based Systems Engineering (MBSE) approach that addresses key shortcomings of traditional object-oriented MBSE tools. 
Principles from the Data-Oriented Programming (DOP) paradigm are used to directly address the deployment complexity, tool integration, and testing difficulties  that have limited adoption of MBSE in aerospace engineering practice.  
This data-oriented MBSE approach separates design data from analysis logic.

The prototype VVERDAD (Vis Viva Engineering, Review, Design,
and Architecture Database) implementation uses the Entity-Component-System (ECS) architecture to represent
heterogeneous engineering data through a generic interface, and treats all data as immutable.
VVERDAD demonstrates that a small set of generic data types can
normalize data from diverse engineering formats (JSON, YAML, TOML, CSV, Excel), enabling format-agnostic template
rendering and analysis integration without requiring engineers to learn a proprietary data model or specialized ontology.

VVERDAD uses the project directory tree as the data namespace, combined with a template system for generating analysis inputs and engineering reports.
This preserves the traceability and reviewability of traditional document-based systems engineering while adding the automation and
consistency checking benefits of a model-based approach. By integrating with standard CI/CD pipelines and containerized analysis tools, the VVERDAD
approach allows discipline engineers to continue using their existing analysis software while participating in an integrated design process. The stateless,
immutable architecture simplifies testing --- each analysis function need only be verified once --- and enables safe parallel
execution without the synchronization complexity inherent in mutable object graphs.
  
The VVERDAD data-oriented MBSE approach is applicable not only to aerospace design but also to mission planning, science planning, mission
operations, and technology development.

\textit{Applicability to Robotic Science Missions:} This approach can not only help to coordinate design work between engineering sub-system teams but also
to better integrate spacecraft design with science planning and instrument design. The data-oriented approach can reduce development
costs, and enable more ambitious, complex science missions and campaigns.

\textit{Applicability to Human Exploration Missions:} For human space exploration, it is vitally important to coordinate the design of multiple
flight elements and to understand the space logistics of operationally aggregating these elements. The generic data interface facilitates
such coordination among flight elements developed across different NASA centers, industry partners, or even via international contributions.

\textit{Applicability to Mission Operations:} This approach can be used not only to design the ground operations elements of a human exploration
or robotic science mission, but also to manage the transfer of design knowledge to the operations teams.

\textit{Applicability to Technology Programs:} Design databases built with this approach can be used to better understand the potential mission applications
for new technologies. Typically design data isn't accessible to technologists unless they are embedded in a project and have the
inside knowledge required to decode the design data.  The data interface makes design data accessible to technologists who can
use it both to demonstrate the potential mission impact of a new technology and to discover unmet mission needs for future research.

\section{Acknowledgement} 

This paper is based upon research supported by the National Aeronautics and Space Administration (NASA) Small Business Innovative Research (SBIR) program
under Contract Number 80NSSC25C0396. Any opinions, findings, and conclusions or recommendations expressed in this material are those of the author and do not 
necessarily reflect the views of the National Aeronautics and Space Administration.

\end{document}